\documentclass{aa}
\usepackage{psfig}
\usepackage{graphicx}

\def\src{PSR J0030+0451}
\def\deg{\ifmmode^{\circ}\else$^{\circ}$\fi} 

\def\lapp{\ifmmode\stackrel{<}{_{\sim}}\else$\stackrel{<}{_{\sim}}$\fi}
\def\gapp{\ifmmode\stackrel{>}{_{\sim}}\else$\stackrel{>}{_{\sim}}$\fi}

\begin{document}

\title{Scintillation measurements of the millisecond pulsar PSR J0030+0451
 and pulsar space velocities}

\author{L. Nicastro \inst{1}
\and F. Nigro \inst{2}
\and N. D'Amico \inst{2}
\and V. Lumiella \inst{3}
\and S. Johnston \inst{4}
}

\institute{
 {Istituto di Fisica Cosmica con Applicazioni all'Informatica, CNR,
 Via U. La Malfa 153, I-90146 Palermo, Italy}
\and 
 {Osservatorio Astronomico di Bologna,
 Via Ranzani 1,  I-40127 Bologna, Italy}
\and 
 {Dipartimento di Astronomia, Universit\`a degli Studi di Bologna,
 Via Ranzani 1,  I-40127 Bologna, Italy}
\and 
 {Research Centre for Theoretical Astrophysics, University of
Sydney, NSW 2006, Australia}
}
\offprints{nicastro@ifcai.pa.cnr.it}
\date{Received 6 December 2000 / Accepted 12 January 2001}

\abstract{
Scintillation observations of the nearby single millisecond pulsar (MSP) \src\
were carried out with the Parkes 64m radiotelescope at three different 
epochs in 1999.
From analysis of the dynamic spectrum we obtained the amplitude of the
electron density power 
spectrum $\log \overline{C_N^2} \simeq -3.33$ and a scintillation velocity 
$V_{\rm iss} \la 15$ km s$^{-1}$. 
This result shows that the Shklovskii effect on the spin-down 
rate $\dot P$ is negligible. We also performed a correlation analysis
between pulsar proper motions ($V_{\rm pm}$) and scintillation velocities
($V_{\rm iss}$) using updated measurements for a sample of 77 objects,
17 of which are MSPs.
The full sample shows a correlation coefficient $r_s \simeq 80\%$ at an
extremely high significance level, while for the
MSP sub-sample (excluding 2 outliers) we obtain $r_s \simeq 90\%$.
\keywords{
Stars: kinematics -- pulsars: individual: \src\ --
 ISM: general -- ISM: kinematics and dynamics --
 ISM: structure -- Radio continuum: stars}
}

\titlerunning{Scintillation measurements of
 PSR J0030+0451 and pulsar space velocities}
\authorrunning{L. Nicastro et al.}

\maketitle

\markboth{L. Nicastro et al.: Scintillation of
  PSR J0030+0451 and pulsar space velocities}
 {L. Nicastro et al.: Scintillation of
  PSR J0030+0451 and pulsar space velocities}

\section{Introduction}
\src\ was independently discovered by the Arecibo Drift Scan Search 
(\cite{somer}) and the Bologna sub-millisecond pulsar survey (\cite{damico}). 
It was the first millisecond pulsar (MSP) discovered with the Northern Cross 
radiotelescope and it appeared to be strongly scintillating at $\lambda=70$ cm.
Follow-up observations performed at Parkes with the 21 cm multibeam receiver
confirmed the pulse period $P \simeq 4.865$ ms,  the
dispersion measure (${\rm DM}) \simeq 4.33$ pc cm$^{-3}$
and showed the pulsar is not in a binary system. Updated ephemeris
from the ongoing monitoring were used
for the data analysis of our scintillation observations.

Given its galactic coordinates $l \simeq 113^{\circ}$,
$b \simeq -57^{\circ}$ and DM,
the Taylor \& Cordes (1993) electron distribution model gives a
distance $D=0.230$ kpc.
Timing observations of \src were carried out by
Lommen et al. (2000) using the Arecibo radiotelescope.
Among the derived parameters, they report
$\dot{P}\simeq (1.0\pm 0.2)\times 10^{-20}$ s s$^{-1}$, and an upper limit
on the pulsar proper motion of $\mu \la (60\div 70)$ mas yr$^{-1}$.
The so called ``Shklovskii effect'' could affect the intrinsic $\dot{P}$
by
\begin{equation}\label{Sl} 
\dot{P}_s/P \simeq 1.1 \times 10^{-22}\;
 V_{\perp}^2 D^{-1}_{\rm kpc}$ s$^{-1},
\end{equation}
where $V_\perp$ is the pulsar transverse velocity in km s$^{-1}$ and 
$D_{\rm kpc}$ its distance in kpc (\cite{shklovskii}; \cite{camilo}).
This puts an upper limit to the proper motion of $\mu \la 60$ mas yr$^{-1}$.

MSPs have an observed mean transverse velocity of $85 \pm 13$ km s$^{-1}$,
a factor $\sim 4$ lower than that of ordinary pulsars,
with those in binary systems having about twice the velocity of the isolated
ones (\cite{toscano1}; \cite{jnk}; \cite{nicastro}). Improving this statistical 
sample is particularly important in the light of neutron star birth scenarios. 
\src's distance, its peculiar position in the Local Interstellar
Medium (LISM) (see \cite{toscano}) and its strong flux variations make 
this MSP a good target for scintillation studies.

The correlation between pulsar proper motions ($V_{\rm pm}$) and
scintillation speeds ($V_{\rm iss}$) was investigated in the past
by several authors (e.g. \cite{hl}; \cite{gupta95}) showing the latter
is a good estimate of the pulsar transverse velocity.
Here we revisit the analysis on a sample of 77 objects, including
\src\ and 16 more MSPs, for which proper motion
(13 are high confidence upper limits)
and scintillation measuremnts exist.
The analysis was performed both using the whole sample and the MSPs subsample
only.

\section{Observations and data analysis}
\begin{figure*}[ht]
\centering                         
{\includegraphics[height=12.5cm,angle=-90]{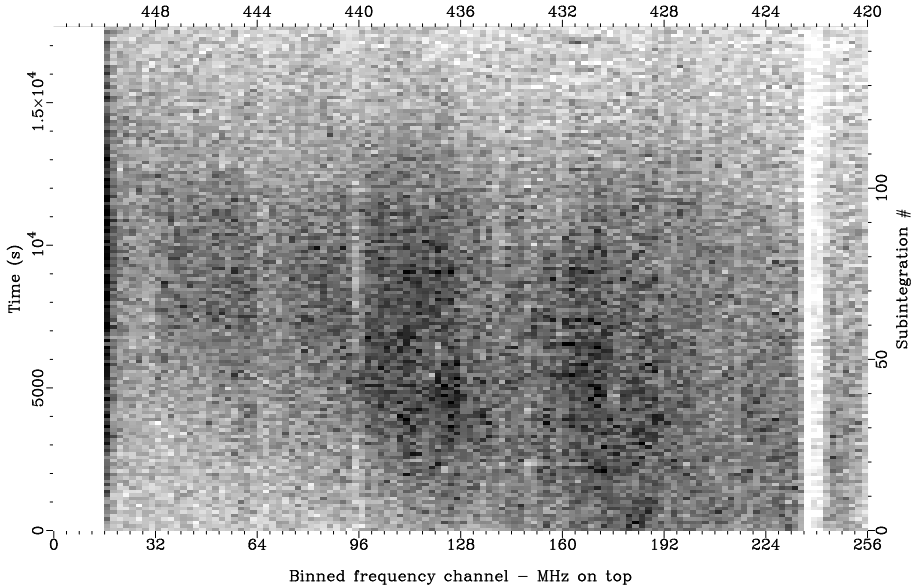}}  
\caption[]{\label{AXP_ctp} Dynamic spectrum of \src\ with a frequency 
resolution of 0.5 MHz and a subintegration time of 960 s. The gray-scale shows
the signal strength increasing from white to black (S/N range $2\div 20$).}
\end{figure*}

During 1999, three observations of \src\ were performed with the 64m Parkes 
radiotelescope at a central frequency of 436 MHz. The total bandwidth was
32 MHz subdivided into 256 filter channels, each of width 125 kHz. The 
filterbanks are sampled continuously at a fixed rate of 0.05 ms; the output
from each channel is one-bit digitized and written to magnetic tape for
off-line analysis. 

The first observation was performed on 1999 July 24 and lasted 7000 s.
The integrated pulsar signal was relatively strong, but unfortunately
the observation time was short compared to the
diffractive scintillation time and only part of the
scintles are visible in the frequency-time dynamic spectrum.
Since this MSP has a northern declination, the longest observation time
from Parkes is about 5.5 hours, and that was the duration of two
more observations performed on October 15 and 16.
In both cases the pulsar signal was weaker than in the first observation
though we were able to detect a number of scintles in the dynamic spectra.

During the analysis, pulse profiles are formed at the apparent pulsar period
over short sections in time and for each of the 256 filter channels.
Subintegration times were 60 s for the first observation and 120 s
for the other two. The signal-to-noise ratios in each subintegration for each 
frequency channel is then calculated. 
The first 16 channels were not considered in the analysis because of their 
malfunctioning; a few other channels that had bad response or in which strong 
interferences were present, were overwritten by an average of the
signal-to-noise 
ratios in adjacent channels. No smoothing operation was made on the dynamic 
spectra but in all the observations a sum along the frequency and time axes
was performed in order to improve the
signal-to-noise ratio. Figure \ref{AXP_ctp} shows a
dynamic spectrum of \src.

To obtain the scintillation bandwidth and decorrelation 
time-scale we performed a two-dimensional autocorrelation analysis on the
dynamical spectra, following the method described by Cordes (1986). 
The scintillation bandwidth, 
$\Delta\nu_{d}$, and the decorrelation time-scale, $t_d$, were obtained
by fitting a Gaussian function to the zero lag in frequency and time 
respectively (details on the data analysis and
description of the software used can be found in \cite{nigro}).
Similar results were obtained with a 2-dimensional gaussian fit.
      
\section{Results and discussion}
Scintillation velocities ($V_{\rm iss}$) can be estimated using the equation
(\cite{gupta94})
\begin{equation}\label{Viss}
V_{\rm iss} = 3.85 \times 10^4\frac{\sqrt{\Delta\nu_d \; D_{\rm kpc} x}}
 {\nu_{\rm GHz} t_d} \; \rm{km\; s}^{-1},
\end{equation}
while the line-of-sight-averaged electron density fluctuations can be
estimated by (\cite{cordes85}; \cite{cordes86})
\begin{equation}
\overline{C_N^2} = 0.002\;\nu_{\rm GHz}^{11/3} D_{\rm kpc}^{-11/6}
 \Delta \nu_d^{-5/6} \; {\rm m}^{-20/3},
\end{equation} 
where $C_{N}^{2}$ is the amplitude of the (density fluctuations) power
spectrum, $\Delta\nu_d$ is the decorrelation bandwidth in MHz, $t_d$ the
characteristic time-scale in s, $D_{\rm kpc}$ the distance in kpc,
$\nu_{\rm GHz}$ the 
observation frequency in GHz and $x\equiv D_{o}/D_{p}$ is the scaling factor
that accounts for an asymmetrically located scattering screen, whith
$D_{o}$ and $D_{p}$ the distances observer--screen and
screen--pulsar, respectively.  
Apart (identifiable) peculiar cases, assuming $x = 1$ and a {\it thin screen}
model (\cite{scheuer}) gives good agreements
between $V_{\rm pm}$ and $V_{\rm iss}$ (see below).
   
\begin{table*}[htb]
\caption[]{\src\ observation log and scintillation parameters.}
\label{tab:1}
\begin{tabular}{lcccccc}
 
\multicolumn{1}{c}{Date} & T & $V_\oplus$ & $\Delta\nu_d$ & $t_d$ &
 $V_{\rm iss}^*$ & $\log \overline{C_N^2}$ \\
 & \multicolumn{1}{c}{(hours)} & (km s$^{-1}$) & (MHz) & (s) & (km s$^{-1}$) &
 (m$^{-20/3}$) \\ \hline
 24--07--1999 & 1.9 & 11.7 & 9.3 & 10569 & 12.5 $\pm$ 11.9 & $-3.66$ \\
 15--10--1999 & 5.5 & 29.2 & 4.9 & 10106 &  9.4 $\pm$  4.0 & $-3.43$ \\
 16--10--1999 & 5.5 & 29.1 & 2.0 &  9518 &  6.4 $\pm$  1.6 & $-3.10$ \\ \hline
\end{tabular}

*~Errors are $1\sigma$ level from equation \ref{errest}.
\end{table*}

Table \ref{tab:1} lists the scintillation parameters of the three observations.
Column 2 to 5 give observation time, transverse Earth velocity toward
the pulsar, decorrelation bandwidth, characteristic time-scale, 
scintillation velocity and the measure of the
turbulence along the line of sight. 
The statistical error on $V_{\rm iss}$ is a combination of two 
quantities: the uncertainties on $\Delta\nu_d$ and $t_d$ from
the gaussian fitting, and the empirical error of the scintillation 
measurements given by Bhat et al. (1998) as
\begin{equation}
\sigma = \left[f_{d}\left(\frac{B_{obs}t_{obs}}
 {\Delta \nu_{d}t_d}\right)\right]^{-0.5},
\label{errest}
\end{equation}
where $\sigma$ is the fractional error, $B_{obs}$ and $t_{obs}$ are the
observation bandwidth
and time respectively, and $f_{d}$ is the filling fraction for number of 
scintles which here is assumed to be 0.5.
This estimate uncertainty is far larger than
that from the fit and that due to the dynamic spectra discretization,
so these two latter are assumed to be negligible.

The scintillation velocity is a combination of (projected on the plane of
the sky perpendicular to the direction of the pulsar)
the pulsar velocity $V_{\rm pm}$, ISM irregularities velocity
$\sigma_{V_{\rm ISM}}$
(assumed to be $\sim 10$ km s$^{-1}$) and Earth orbital velocity $V_\oplus$
(varying from 0 to $\sim 30$ km s$^{-1}$). In the case of a binary system,
the pulsar's orbital velocity should also be considered.
For pulsars with high values of $V_{\rm iss}$, $\sigma_{V_{\rm ISM}}$ and $V_\oplus$
give a negligible contribution to the estimate of the transverse velocity
(see e.g. \cite{gupta94}).
For \src\ the values of $V_{\rm iss}$ are similar to $\sigma_{V_{\rm ISM}}$
and in two, out of three cases, it is $V_{\rm iss} < V_\oplus$.
Therefore these velocities contribute significantly to the uncertainty.
A possible way to consider these systematics is by
\begin{equation}
\sigma_{V_{\perp}}=\sigma_{V_{\rm iss}} +
 \frac{\sqrt{\sigma_{V_{\rm ISM}}^2+V_\oplus^2}}{2}. 
\end{equation}
However, to better estimate the velocity $V_{\perp}$ of
the pulsar and the related uncertainty,
a more reliable way would be performing a
campaign of measurements preferably when the contribution of $V_\oplus$
is minimum ($\sim 5$ km s$^{-1}$). In addition, having several measurements
would allow to assume $\overline{V_{\rm ISM}}= 0$. We have planned to do so.
Neverthless our measurements clearly indicate \src\ is a slowly moving object.

\subsection{LISM}
The average amplitude of the electron density power spectrum value
$\log \overline{C_N^2} = -3.33$ we found for \src\ is somewhat greater than
$-3.5$, the average
value for the whole pulsar sample with distances $\ga 1$ kpc, i.e. out of
the LISM (\cite{cordes86}).
This estimate adds to that of other $\sim 50$ pulsars with
DM $ \la 35$ pc cm$^{-3}$ (20 by Bhat et al. (1998) and 31 by
Johnston et al. (1998)).
Most of these pulsars are localized within the LISM and have high
$\overline{C_N^2}$ values, confirming
the existence of local enhanced scattering.
The overall picture 
shows clearly that the LISM is highly turbulent compared to the outer regions.
Bhat et al. (1998) developed a model which they claim can fit the
$\overline{C_N^2}$ observed in their pulsar sample.
It is interesting to check their model against \src\ results
since it is located in the quadrant where no other MSPs are
known (\cite{toscano}).

\begin{figure}[htb]
\centering
{\includegraphics[width=8.5cm,angle=0]{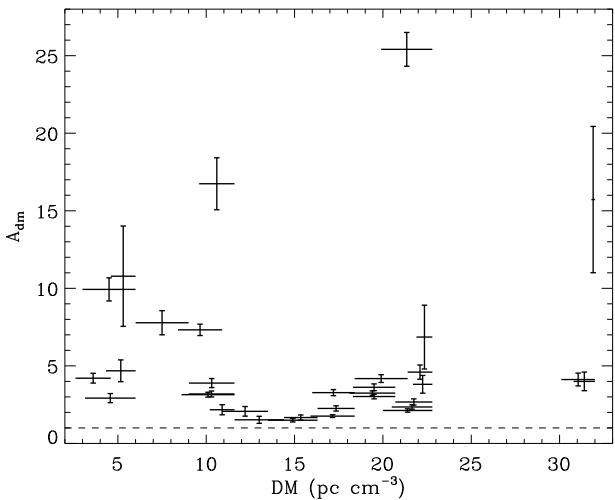}}
\caption[]{\label{fig:adm} ``Anomaly factor'' vs DM for 35 PSRs with
 ${\rm DM <35}$ pc cm$^{-3}$. The 3 isolated highest
  points are due to PSRs J1730$-$2304, J1825$-$0935 and J2129$-$5718.}
\end{figure}

The expected decorrelation bandwidth value, given by
\begin{equation}
\Delta \nu_{d,t} \simeq 11\nu^{22/5}_{\rm GHz}D^{-11/5}_{\rm kpc} = 7.2
 \; {\rm MHz},  
\end{equation}
is not too different from the mean measured value $\Delta\nu_d = 5.6$ MHz.
We computed the so called ``anomaly parameter'' (defined as
$A_{\rm dm}=(\nu_{d_1}/\nu_{d_2})_{\rm obs}/(\nu_{d_1}/\nu_{d_2})_{\rm exp}$,
see \cite{bhat98}) for \src\ with respect to
PSR J1133+16 and PSR J1929+10, from the Bhat et al. (1998) sample, and
PSR J1744$-$1134 and J2124$-$3358 (both MSPs) from the Johnston et al. (1998)
sample, as they all have similar distances and, more importantly, DMs
within $\simeq 1$ pc cm$^{-3}$. The low values of $A_{\rm dm}$ we obtain
(1.4, 2.4, 1.60 and 1.34 respectively) do not confirm the decay trend
shown in Figure 7 of Bhat et al. (1998).
An asymmetry in the distribution
of electrons in the LISM (as suggested by \cite{toscano}) or a non-Kolmogorov
density spectrum could explain the discrepancy.
This is also shown in Fig. \ref{fig:adm} where we plot
$A_{\rm dm}$ vs DM for 35 PSRs (16 MSPs) with DM $<35$ pc cm$^{-3}$. This
sample includes 32 objects from Johnston et al. (1998) and the 3 MSPs:
\src, J1012+5307 and B1257+12.
The plot was made by averaging the $A_{\rm dm}$ of pulsars with DM differences
less than 3 pc cm$^{-3}$ and increasing the DM by that of the pulsars instead of
a fixed step of 1 pc cm$^{-3}$ (as is done by \cite{bhat98}).
Error bars are inversely proportional to the number of pulsar products averaged.
The three very high peaks at DMs 10, 21 and 32 pc cm$^{-3}$ are due
to the large discrepancies between the
expected decorrelation bandwidths and the measured ones for
PSRs J1730$-$2304, J1825$-$0935 and J2129$-$5718, respectively.
The estimators Bhat et al. (1998) use to define the best model for the
local scattering structure are in our case
$\epsilon_A({\rm DM})=0.119$ and $\epsilon_b=0.275$, confirming
the inhomogeneity of the LISM.

However the high variability of the observed decorrelation bandwidth in \src\
by a factor $\sim 5$ is rather unusual. This is not caused by
refractive scintillation because the dynamic spectra show no frequency drift.
Instead it likely originates from anisotropies in the density or the
magnetic field structure of the LISM.
Further and more sensitive observations are required
to put any strong constraint on the proposed model for the LISM
by Bhat et al. (1998).

\subsection{Transverse velocity}
The scintillation parameters allow us to derive a reliable estimate of the
transverse velocity of \src. We will show that the correlation between
$V_{\rm iss}$ and $V_{\rm pm}$ is very high and a relation
$V_{\rm iss}\equiv V_{\rm pm}$ can in general be assumed
(see also \cite{gupta95}).
As already mentioned, in spite of an average
high flux density measured during the first observation, because of its short
duration compared to $t_d$, the $V_{\rm iss}$ estimate is affected by a
relatively large error.
Further, the data of the second and third observation
have lower statistical significance, so they can provide only upper limits
(see also Tab. \ref{tab:1}). This suggests to give similar weights to the
three observations in the calculation of $\overline{V_{\rm iss}}$.
Hence we have determined a mean velocity of
$\overline{V_\perp} \simeq 9\pm 6$ km s$^{-1}$ ($1\sigma$ error).
This number makes \src\ the pulsar with the smallest 
transverse velocity so far measured.
Its position in the $V_{\rm pm}$ vs. $V_{\rm iss}$ scatter plot is
quite isolated in the bottom left corner well in agreement with a correlation
$V_{\rm pm}\equiv V_{\rm iss}$ (see Fig. \ref{fig:vpm_viss}).
 
If we assume that the $\dot{P}=(1\pm 0.2) \times 10^{-20}$
reported by Lommen et al. (2000) is all due to
the Shklovskii effect, we obtain an upper limit to the
transverse speed $V_\perp< 65$ km s$^{-1}$.
Now assuming $\overline{V_\perp}=9$ km s$^{-1}$,
the derived contribution to the spin-down rate is
$\dot{P}_s < 2 \times 10^{-22}$ s s$^{-1}$.
This means that the Shklovskii effect introduces an error
of only $\sim 2\%$ on $\dot{P}=\dot{P}_i+\dot{P}_s$, lower than the $\sim20\%$
so far obtained by the timing data fit. Hence we can assume the 
measured spin-down rate is the intrinsic one: $\dot{P}_i = (1.0 \pm 0.2) 
\times 10^{-20}$ s s$^{-1}$ (\cite{lommen}).
However, using a $3\sigma$ upper limit
$\overline{V_\perp} \simeq 27$ km s$^{-1}$, we obtain a $\dot{P}_s$
comparable to the $\dot{P}$ error estimate. We then conclude that
$\dot{P}_i \cong (0.8 \div 1) \times 10^{-20}$ s s$^{-1}$, giving a
surface magnetic field $B \cong (2\div 2.2) \times 10^{8}$ G and
a characteristic age $\tau \cong (8\div 10) \times 10^{9}$ years.

This pulsar was also detected in the soft X-ray band 
(\cite{becker}). Its X-ray luminosity (0.1--2.4 keV) is
$L_x \sim (1\div 2) \times 10^{30}(d/0.23\; {\rm kpc})^2$ erg s$^{-1}$.
Using the inferred rotational energy loss rate 
$\dot{E}= (2.7 \div 3.4) \times 10^{33}$ erg s$^{-1}$
we obtain a relatively narrow X-ray efficiency range $\eta = L_x/\dot{E}$ 
$\simeq (0.3\div 0.7) \times 10^{-3} (d/0.23\; {\rm kpc})^2$. 
This is in good agreement with the proposed law
$L_x/\dot{E} \simeq 10^{-3}$ (\cite{bt}) and
points toward a magnetospheric origin for the high energy radiation, in spite
the inferred rotational energy loss rate and pulse period 
would put \src\ in the class of MSPs for which the 
X-ray emission is believed to be mostly thermal
(\cite{ks}).
Becker et al. (2000), from the observed X-ray pulse profile
(very similar to the radio one) and the large pulsed fraction 
($\sim 69 \pm 18\%$) also suggest a non-thermal origin of the X-ray 
emission. Thanks to the Chandra and XMM-Newton X-ray satellites,
a good quality X-ray spectrum should soon be able to give a
definitive answer in this respect.

\src\ is the 8$^{th}$ isolated MSP detected for which a transverse velocity
is available.
Using for \src\ $V_{\rm pm} = 32$ km s$^{-1}$ (i.e. $\simeq 65/2$),
the $\overline{V_{\rm pm}}$ for 7 MSPs (we exclude PSR J1730$-$2304)
is $58 \pm 6$ km s$^{-1}$ (for 5 MSPs $\overline{V_{\rm iss}} \simeq 45$
 km s$^{-1}$).
This, even if at low statistical significance, is nearly a factor 2 lower 
than the $96\pm 9$ km s$^{-1}$ found for 18 binary MSPs
(for 12 MSPs $\overline{V_{\rm iss}} \simeq 113$ km s$^{-1}$)
(\cite{jnk}; \cite{toscano1}).
 
\section{$V_{\rm pm}$--$V_{\rm iss}$ correlation}
\begin{figure*}[htb]
\centering                         
{\includegraphics[width=16cm,angle=0]{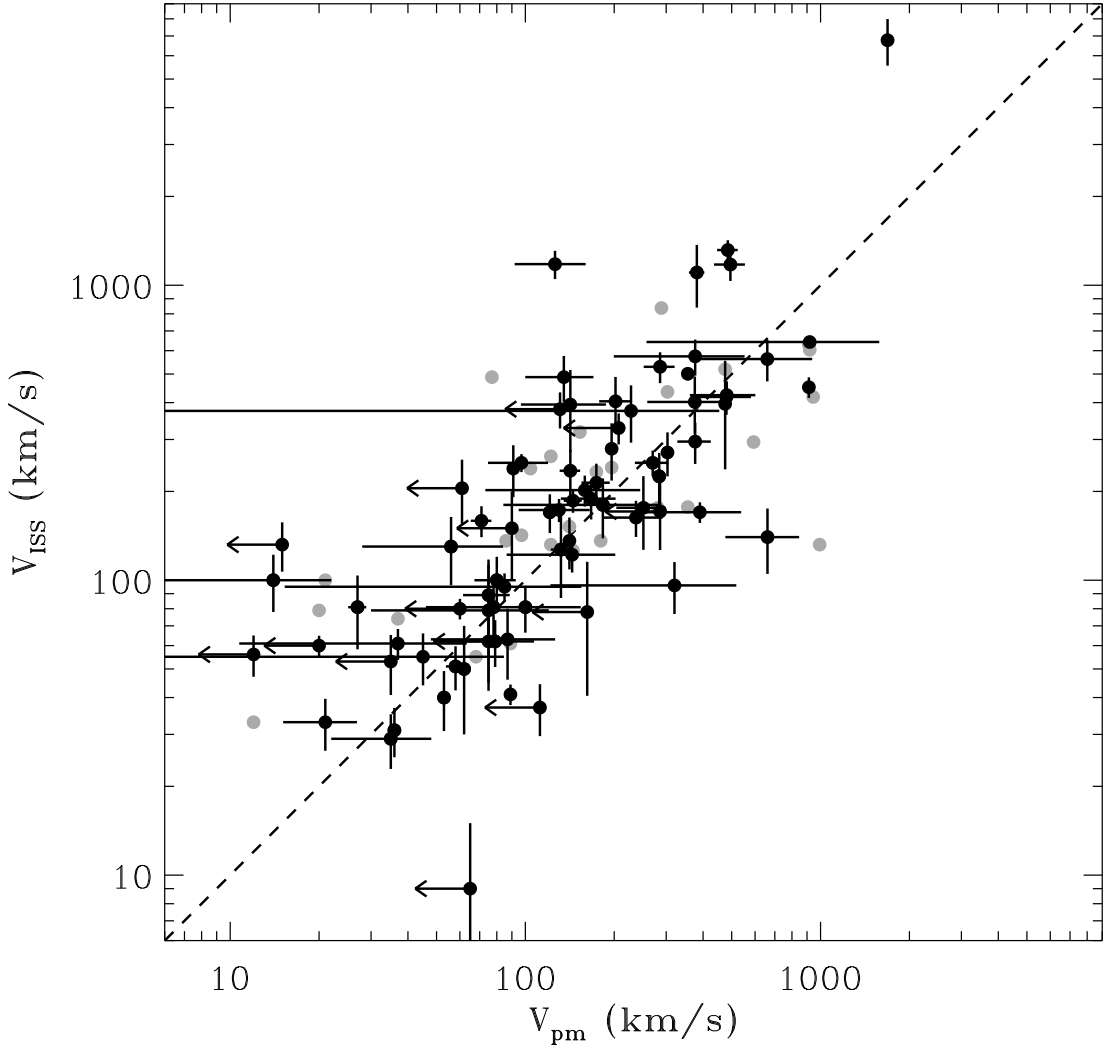}}  
\caption[]{\label{fig:vpm_viss} Scatter diagram for proper motion and
 scintillation speeds for 77 pulsars.
 The grey points are the pulsars with updated values compared to Gupta (1995).
 The $V_{\rm pm}\equiv V_{\rm iss}$ correlation line is shown.}
\end{figure*}

%
\begin{table*}[t]
\caption[]{Proper motion and scintillation velocities for 77 pulsars.}
\label{tab:vpm_viss}
\begin{center}
\begin{tabular}{lrrrrlrrrr}

\multicolumn{1}{c}{PSR} & \multicolumn{1}{c}{$\overline{V_{\rm pm}}$} &
 \multicolumn{1}{c}{$\delta\overline{V_{\rm pm}}$} &
 \multicolumn{1}{c}{$\overline{V_{\rm iss}}$} &
 \multicolumn{1}{c}{$\delta\overline{V_{\rm iss}}$} &
\multicolumn{1}{c}{~PSR} & \multicolumn{1}{c}{$\overline{V_{\rm pm}}$} &
 \multicolumn{1}{c}{$\delta\overline{V_{\rm pm}}$} &
 \multicolumn{1}{c}{$\overline{V_{\rm iss}}$} &
 \multicolumn{1}{c}{$\delta\overline{V_{\rm iss}}$} \\
\multicolumn{1}{c}{~(J2000)} &\multicolumn{1}{c}{(km s$^{-1}$)} & \multicolumn{1}{c}{(\%)} &
 \multicolumn{1}{c}{(km s$^{-1}$)} & \multicolumn{1}{c}{(\%)}    &
\multicolumn{1}{c}{(J2000)} &\multicolumn{1}{c}{(km s$^{-1}$)} & \multicolumn{1}{c}{(\%)} &
 \multicolumn{1}{c}{(km s$^{-1}$)} & \multicolumn{1}{c}{(\%)}    \\
\hline
0304+1932  & 167  &  21  &   189   &    15  & ~~~1907+4002  & 126  &  27  & 1179   &    11  \\
0323+3944  & 237  &  23  &   163   &    14  & ~~~1913$-$0440  & $<372$  &      &  380   &    14  \\
0332+5434  & 145  &   7  &   186*  &     9  & ~~~1921+2153  & 130*  &  27  &  173*  &     9  \\
0358+5413  & 135* &  26  &   488   &    18  & ~~~1932+1059  &  71* &   8  &  159*  &    12  \\
0452$-$1759& 320*  &  62  &    96*  &    20  & ~~~1935+1616  & 480*  &  21  &  418   &    13  \\
0454+5543  & 202  &  12  &   404   &    21  & ~~~1939+2134  &  $<146$  &      &  205   &    25  \\
0528+2200  & 228  &  99  &   375   &    22  & ~~~1946+1805  &  37  &  71  &   61*   &    12  \\
0543+2329  & 376  &  47  &   574   &    14  & ~~~1954+2923  &  87  &  45  &   63   &    27  \\
0614+2229  & $<304$  &      &    37   &    20  & ~~~1955+5059  & 495  &  12  & 1175   &    12  \\
0629+2415  & $<538$  &      &    78   &    48  & ~~~2018+2839  &  $<31$  &      &   56*  &    16  \\
0630$-$2834& 390* &  38  &   170*  &     8  & ~~~2022+2854  &  97  &  23  &  250*  &     7  \\
0659+1414  & 251* &  19  &   176   &    28  & ~~~2022+5154  &  91* &   2  &  239   &    20  \\
0814+7429  &  75  &  18  &    89   &    32  & ~~~2046$-$0421  & $<532$  &      &  328   &    12  \\
0820$-$1350& 376  &  13  &   295   &    16  & ~~~2046+1540  & 159  &  54  &  202   &    12  \\
0823+0159  &  $<127$  &      &    53   &    23  & ~~~2048$-$1616  & 355* &   4  &  501*  &     5  \\
0826+2637  & 196  &  3.3 &   279*  &    22  & ~~~2113+2754  & 381  &   6  & 1105   &    24  \\
0835$-$4510& 141  &   4  &   136*  &    20  & ~~~2116+1414  & $<618$  &      &  171   &    26  \\
0837+0610  & 174  &  11  &   214*  &    16  & ~~~2157+4017  & 485  &   8  & 1316   &     8  \\
0908$-$1739& 142  &  32  &   394   &    31  & ~~~2219+4754  & 375  &  31  &  402   &    22  \\
0922+0638  & 919  &  72  &   642*  &     4  & ~~~2225+6535 & 1686  &  2.3 & 6772   &    18  \\
0953+0755  &  21  &  28  &    33*  &    20  & ~~~2305+3100  & 661  &  28  &  140   &    25  \\
1115+5030  & 142  &   8  &   235   &    18  & ~~~0030+0451$\dagger$  &  32  & 100  &    9   &    66  \\
1136+1551  & 475  &  1.6 &  396*  &    40  & ~~~0437$-$4715  & 121   &  0.2 &  170   &    15   \\
1239+2453  & 303  &  4.4 &  271*  &    17  & ~~~0711$-$6830$\dagger$  &  78   &  2.5 &   81   &    20  \\
1456$-$6843&  89  &  3.4 &   41*   &     8  & ~~~1012+5307   & 62$^{1}$  & 0.5  &  50 &  40 \\
1509+5531  & 913  &  5   &  451*  &     8  & ~~~B1257+12  &  284   &  1   &   225$^2$  &    20  \\
1543$-$0620&  $<170$*  &      &   80*  &     8  & ~~~1455$-$3330  & 100   & 54   &   81   &    18  \\
1543+0929  & 144  &  40  &  122   &    13  & ~~~B1534+12  &  132   &   8   &   161   &    20  \\
1607$-$0032  &  $<66$  &      &   60*  &     8  & ~~~1603$-$7202  &  27   &  7   &   81   &    28  \\
1645$-$0317  & 660  &  42  &  562   &    16  & ~~~1713+0747  &  35$^{1}$   &  3.2 &   29   &    21  \\
1709$-$1640  &  $<70$* &      &  132   &    19  & ~~~1730$-$2304$\dagger$  &  79  &  35   &   62   &    18  \\
1752$-$2806  &  45*  &  88  &   55   &    20  & ~~~1744$-$1134$\dagger$  &  36$^1$  &   1.8 &   31   &    19  \\
1807$-$0847  &  85*  &  82  &   95*  &    11  & ~~~1911$-$1114  & 183  &  54   &  180   &    23  \\
1820$-$0427  & 270  &  13  &  250   &    10  & ~~~2051$-$0827  &  14  &  57   &  100   &    22  \\
1823+0550  &  $<238$  &      &   62   &    32  & ~~~2124$-$3358$\dagger$  &  53  &   3.8 &   40   &    23  \\
1825$-$0935  &  75* &  60  &   79*  &    43  & ~~~2129$-$5718  &  56  &  50   &  130   &    26  \\
1840+5640  & 286  &  12  &  529   &    12  & ~~~2145$-$0750  &  58$^1$  &  11   &   51   &    17   \\
1844+1454  & 481  &  25  &  424   &    11  & ~~~2317+1439  &  80  &  16   &  100   &    20  \\
\hline
\multicolumn{10}{l}{*~Updated velocity estimates respect to Gupta (1995).
 $\dagger$~Isolated MSP.} \\
\multicolumn{10}{l}{$^1$~Lange et al. (2000).
  $^2$~Gothoskar \& Gupta (2000).}
 
\end{tabular}
\end{center}
\end{table*}
Gupta (1995) made a comparison between proper motion and scintillation
velocity for 59 pulsars using the Taylor \& Cordes (1993) model to calculate
their distances and a new formula to calculate the ISS velocities
(\cite{gupta94}).
The comparison showed a good correlation between $V_{\rm pm}$ and $V_{\rm iss}$. 
The correlation coefficient he found is $r=0.42$. 
It becomes 0.47 if pulsars for which only upper limits were available are 
excluded from the sample. A least square fit gave a slope of $1.1\pm 0.13$.

We updated the $V_{\rm pm}$--$V_{\rm iss}$ diagram
\begin{enumerate}
\item adding the proper motion and scintillation speeds of 16 MSPs
 (\cite{toscano1}; \cite{lange}; \cite{jnk}; \cite{gg}; \cite{lorimer};
 this paper) and of the relativistic pulsar B1534+12 (\cite{stairs}),
\item using 14 revised proper motion estimates (\cite{fo97}; \cite{fo99}),
\item using revised values of $V_{\rm iss}$ for 22 pulsars
 (\cite{bhat99}; \cite{lumiella}), plus PSR B1259$-$63 (\cite{mcg}).
\end{enumerate}
The $V_{\rm pm}$ vs. $V_{\rm iss}$ values for 77 pulsars are shown in
the scatter 
plot of Fig. \ref{fig:vpm_viss} and listed in Tab. \ref{tab:vpm_viss}, (MSPs
are listed at the bottom of the Table). The uncertainties on $V_{\rm iss}$ were 
computed according to Bhat el al. (1998) except for
PSR B1257+12 and B1534+12 for 
which we gave a 20\% error. For the uncertainties on $V_{\rm pm}$ we took
into account the proper motion errors only.
We know that uncertainties
in the distance from the Taylor \& Cordes model are on average $\sim 30\%$
but we do not include this source of error.
Only in a few cases do distances come from parallax measurements.

For \src\ we adopted a (low statistical significance) proper motion value
of $V_{\rm pm}=32 \pm 32$ km s$^{-1}$ (from the Shklovskii upper limit)
and $V_{\rm iss}= 9\pm 6$  km s$^{-1}$ (that is the average of the values of
Tab. \ref{tab:1}). 
The object at the top of the plot is PSR J2225+6535.
Toscano et al. (1999a) gave for PSR J1730$-$2304 a lower limit of
$V_{\rm pm}\ga 51$ km s$^{-1}$ and an upper limit of
$V_{\rm pm}\la 107$ km s$^{-1}$ from the
Shklovskii effect.
We averaged these value and used the estimate of $79 \pm 28$ km s$^{-1}$.
For the relativistic pulsar PSR B1534+12 we used the inferred distance
$D = 1.1$ kpc obtained by Stairs et al. (1998) assuming the validity of general 
relativity theory.

A correlation analysis gives for the total sample a coefficient $r = 75\%$, 
much higher than the value obtained by Gupta (1995). Further statistical 
analysis, using a rank Spearman test, gives a correlation coefficient
$r_s = 79\%$ at an extremely significance level of $\sim 1\times 10^{-17}$
(the lower the value the higher the confidence of the result
(\cite{numerical})). A least squares fit with a linear law gives a slope
of $0.81\pm 0.21$. Changing the axes of the scatter plot  
(therefore taking into account the $V_{\rm pm}$ error values) we obtain
a slope of $0.71\pm 0.20$.
Excluding the 14 points with upper limits on $V_{\rm pm}$,
we obtain $r =78\%$ and 
$r_s = 80\%$ at a significance level of $\sim 1\times 10^{-15}$ and a
slope of $0.88\pm 0.24$.
Exchanging the axes of the plot the least squares fit gives an angular
coefficient of $0.70\pm 0.25$.
So we may estimate the slope of a straight line that fits the points
taking into account the errors in both axis as $m \simeq 
0.90 \pm 0.21$ for the whole sample, and $m \simeq 0.82 \pm 0.25$ excluding
the upper limits.

\subsection{MSP correlation}
PSRs J1603$-$7202 and J2051$-$0827 are the MSPs for which
$V_{\rm iss}$ differs significantly from $V_{\rm pm}$.
To have $V_{\rm pm}\equiv V_{\rm iss}$ the scaling factor should be
$x\simeq 3.0$ and $x\simeq 7.1$, respectively.
These values suggest the presence 
of a scattering medium near the neutron star.
This result is in agreement with the theory that the binary
system containing PSR J1603$-$7202 evolved through a phase with critical
unstable mass-transfer and most likely hosts a CO--WD companion rather than 
a He--WD companion (\cite{tauris}). Binary systems like this are believed
to be surrounded by 
an envelope that slowly expands giving rise to the mass transfer (\cite{van}).
This envelope is thought to be the main cause of scintillation.
For the eclipsing binary pulsar PSR J2051$-$0827, the scintillation 
parameters vary significantly with orbital phase (\cite{sta}; \cite{jnk}).
It is likely that the wind blown from the companion star is contributing 
to the scintillation parameters.
In these cases, because the screen is so close to the pulsar,
the measured scintillation velocity is effectively the velocity of the
screen (see e.g. equation 3 in Britton et al. 1998). In both cases,
the wind could easily have a velocity of $\sim$100 km s$^{-1}$.
One possible test of this idea would be to perform VLBI observations
of these two pulsars to resolve (or not) the scattering disk; such
an observation would give a direct measurement of the screen distance.

\begin{figure}[htb]
\centering
{\includegraphics[width=8.5cm,angle=0]{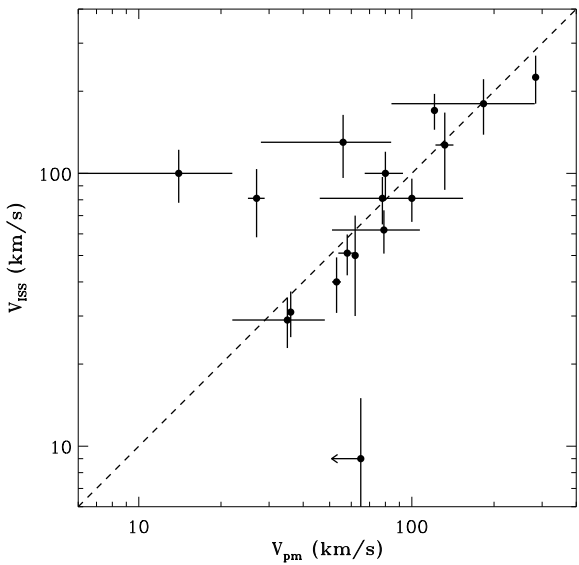}}  
\caption[]{\label{MSPvpm_viss} Scatter diagram for proper motion and 
 scintillation speeds for 16 millisecond pulsars.
 The $V_{\rm pm}\equiv V_{\rm iss}$ correlation line is shown.}
\end{figure}

We performed a correlation analysis between $V_{\rm pm}$ and $V_{\rm iss}$
for the sub-sample of 16 MSPs plus PSR B1534+12.
Figure \ref{MSPvpm_viss} shows the scatter plot.
A simple correlation analysis gives a coefficient $r = 60\%$.
A rank Spearman test 
gives a correlation coefficient $r_s = 67\%$ at a significance level
of $\sim 3 \times 10^{-3}$. A least squares fit with a linear law gives a 
slope of $\sim 0.57$. Exchanging the axis of the scatter plot the least squares
fit gives a slope of $\sim 0.47$.
Excluding the two above mentioned MSPs from the analysis, 
the correlation coefficient becomes $r = 87\%$
and $r_s$ = 91\% at a much higher significance level of $\sim 3\times 10^{-6}$. 
A least squares fit with a linear law gives a slope of $\sim 1.1$.
Exchanging the axis of the plot the slope becomes $\sim 0.5$.
So we estimate the slope of the linear fit to be
$m \sim 1.1$ for the whole sample, and $m \sim 1.6$ 
excluding PSR J1603$-$7202 and J2051$-$0827.

Since MSPs have low speeds, a rigorous analysis should take properly
into consideration systematical uncertainties due to the Earth orbital
velocity and the ISM irregular motion.
In spite of that the resulting $V_{\rm pm}$--$V_{\rm iss}$ correlation is
excellent. This demonstrates that averaging several observations spread over an
Earth orbit is (in general) sufficient to wash out systematics.
 
\section{Conclusions}
We presented observational and derived parameters for 
the diffractive scintillation of the recently discovered \src.
Our results for this source show that:
\begin{itemize}
 \item its decorrelation bandwidth is highly variable;
 \item the Kolmogorov coefficient $\log \overline{C_N^2}=-3.33$ along
 its line of sight is somewhat
 higher than the average for pulsars out of the LISM where it is
 $\log \overline{C_N^2}=-3.5$.
 The enhanced turbulence and strong changes in $\nu_d$ support the findings
 of Bhat et al. (1998) about the
 LISM, but seems to not be in agreement with their proposed model for the
 electron distribution;
 \item it is unlikely its $V_\perp$ exceeds 30 km s$^{-1}$;
 \item as a consequence of this, the contribution of $V_\perp$ to the
 observed spin-down rate through the Shklovskii effect is negligible and
 suggests the derived parameters $B$, $\tau$ and $\dot E$ are not affected
 by spurious effects. In particular
 $\dot E = (2.7 \div 3.4) \times 10^{33}$ erg s$^{-1}$ and the observed
 X-ray luminosity are fully compatible with an X-ray efficiency factor
 $\eta = 1\times 10^{-3}$. However the low observed $L_x$ suggests this
 pulsar is somehow in between the two proposed
 classes of X-ray emitting pulsars (thermal polar caps and magnetospheric
 emission).
 Only high quality X-ray spectral data will allow to discriminate between
 these two emission mechanisms. The X-ray new generation 
 observatories XMM-Newton and Chandra can provide these data.
\end{itemize}
To obtain accurate estimates of the scintillation parameters,
very long observation times and large bandwidths are needed.
Because of its high sensitivity and site location in the North,
the Indian Giant Metre Radio Telescope (GMRT) is the most appropriate
instrument for \src\ scintillation studies.
For minimizing systematical errors, the 
observations must preferably be carried out at epochs when the
projected Earth transverse orbital speed is low.   
Within a few years we should also be able to have an accurate timing
measurement of $V_{\rm pm}$,
although our results already show that \src\ is one of the
slowest pulsars (with known $V_\perp$).

The pulsar sample for which proper motion and scintillation 
speeds are available was revised with new values for ordinary pulsars and
17 (recycled) millisecond pulsars. We performed several different correlation
analysis on the resulting sample of 77 objects and found:
\begin{itemize}
 \item assuming the very simple model with the scattering screen placed
 midway between the Earth and the pulsar, the overall
 $V_{\rm pm}$--$V_{\rm iss}$ correlation
 is $r\simeq 80\%$; this value is much higher than the one reported by
 Gupta (1995);
 \item MSPs show a good correlation $r_s\simeq 65\%$ that becomes
 $\simeq 88\%$ if two anomolous PSRs (J1603$-$7202 and J2051$-$0827),
 for which {\em local} enhanced
 electron densities were reported, are excluded. This confirms that
 interstellar scintillation parameters can be used as
 a probe of enhanced electron densities along the line of sight to PSRs;
 \item we confirm that multi-epoch scintillation observations
 wash out systematics and give
 reliable estimates of the pulsar transverse velocity.
\end{itemize}

\begin{acknowledgements}
F.N. wish to thank IRA-CNR for technical support.
The Australia Telescope is funded by the Commonwealth of Australia
for operation as a National Facility managed by the CSIRO.
\end{acknowledgements}


\begin{thebibliography}{}

\bibitem[Becker \& Tr\"umper\ 1997]{bt}
 Becker W., Tr\"umper J., 1997, A\&A 326, 682

\bibitem[Becker et al.\ 2000]{becker}
 Becker W., Tr\"umper J., Lommen A.N., Backer D.C., 2000, ApJ 545, 1015

\bibitem[Bhat et al.\ 1998]{bhat98}
 Bhat N.D.R., Gupta Y., Rao A.P., 1998, 
 ApJ 500, 262

\bibitem[Bhat et al.\ 1999]{bhat99}
 Bhat N.D.R., Rao A.P., Gupta Y., 1999, 
 ApJS 121, 483

\bibitem[Britton et al.\ 1998]{brit98}
 Britton M.C., Gwinn C.R., Ojeda M.J., 1998,
 ApJ 501, L101

\bibitem[Camilo et al.\ 1994]{camilo}
 Camilo F., Thorsett S.E., Kulkarni S.R., 1994, 
 ApJ 421, L15

\bibitem[Cordes et al.\ 1985]{cordes85}
 Cordes J.M., Weisberg J.M., Boriakoff V., 1985, 
 ApJ 288, 221

\bibitem[Cordes\ 1986]{cordes86}
 Cordes J.M., 1986, 
 ApJ 311, 183

\bibitem[D'Amico\ 2000]{damico}
 D'Amico N., 2000, In: Pulsar Astronomy - 2000 and Beyond, 202th 
 ASP Conf. Ser., M. Kramer, N. Wex \& R. Wielebinski (eds.), 
 San Francisco: ASP, p.27 
 
\bibitem[Fomalont et al.\ 1997]{fo97} 
 Fomalont E.B., Goss W.M., Manchester R.N., Lyne A.G., 1997, 
 MNRAS 286, 81

\bibitem[Fomalont et al.\ 1999]{fo99}
 Fomalont E.B., Goss W.M., Beasley A.J., Chatterjee S., 1999, 
 ApJ 117, 3025

\bibitem[Gothoskar \& Gupta\ 2000]{gg}
 Gothoskar P., Gupta Y., 2000, 
 ApJ 531, 345

\bibitem[Gupta et al.\ 1994]{gupta94}
 Gupta Y., Rickett B.J., Lyne, A.G., 1994, 
 MNRAS 269, 1035

\bibitem[Gupta\ 1995]{gupta95}
 Gupta Y., 1995, 
 ApJ 451, 717

\bibitem[Harrison \& Lyne\ 1993]{hl}
 Harrison P.A., Lyne A.G., 1993, MNRAS 265, 778

\bibitem[Johnston et al.\ 1998]{jnk}
 Johnston S., Nicastro L., Koribalski B., 1998, 
 MNRAS 297, 108

\bibitem[Kawai \& Saito\ 1999]{ks}
 Kawai N., Saito Y., 1999, 
 Astro. Lett. and Communications 38, 1

\bibitem[Lange et al.\ 2000]{lange}
 Lange Ch., Wex N., Kramer M., Doroshenko O., Backer D.C., 2000, 
 In: Pulsar Astronomy - 2000 and Beyond. 202th 
 ASP Conf. Ser., M. Kramer, N. Wex \& R. Wielebinski (eds.), 
 San Francisco: ASP, p.61

\bibitem[Lommen et al.\ 2000]{lommen}
 Lommen A.N., Zepka A., Backer D.C., McLaughlin M., Cordes J.M.,
 Arzoumanian Z., Xilouris K., 2000, ApJ 545, 1007

\bibitem[Lumiella\ 2000]{lumiella}
 Lumiella V., 2000, 
 Tesi di Laurea, Universit\`a degli Studi di Bologna, Italy

\bibitem[Lorimer et al.\ 1995]{lorimer}
 Lorimer D.R., Lyne A.G., Festin L., Nicastro L., 1995, 
 Nat 376, 393

\bibitem[McClure-Griffiths et al.\ 1998]{mcg}
 McClure-Griffiths N. M., Johnston S, Stinebring D.R., Nicastro L, 1998,
 MNRAS 492, L49

\bibitem[Nicastro \& Johnston\ 1995]{nicastro}
 Nicastro L., Johnston S., 1995, 
 MNRAS 273, 122

\bibitem[Nigro\ 2000]{nigro}
 Nigro F., 2000,
 Tesi di Laurea, Universit\`a degli Studi di Bologna, Italy

\bibitem[Press et al.\ 1992]{numerical}
 Press W., Teukolsky S., Vetterling W., Flannery B., 1992,
 Numerical recipes in Fortran: The art of scientific computing (2nd edition), 
 Cambridge University Press, p.633

\bibitem[Scheuer\ 1968]{scheuer}
 Scheuer P.A.G., 1968, 
 Nat 218, 920

\bibitem[Shklovskii\ 1970]{shklovskii}
 Shklovskii I.S., 1970, 
 Soviet Astron. 13, 562

\bibitem[Somer\ 2000]{somer}
 Somer A., 2000, In: Pulsar Astronomy - 2000 and Beyond, 202th 
 ASP Conf. Ser., M. Kramer, N. Wex \& R. Wielebinski (eds.), 
 San Francisco: ASP, p.17

\bibitem[Stairs et al.\ 1998]{stairs}
 Stairs I.H., Arzoumanian Z., Camilo F., Lyne A.G., Nice D.J., Taylor J.H.,
 Thorsett S.E., Wolszczan A., 1998, 
 ApJ 505, 352

\bibitem[Stappers et al.\ 1996]{sta}
 Stappers B.W., Bailes M., Lyne A.G., Manchester R.N., D'Amico N., Tauris T.M.,
 Lorimer D.C., Johnston S., Sandhu J.S., 1996, 
 ApJ 465, L119

\bibitem[Tauris \& Savonije\ 1999]{tauris}
 Tauris T.M., Savonije G.J., 1999, 
 A\&A 350, 928

\bibitem[Taylor \&  Cordes\ 1993]{tc}
 Taylor J.H., Cordes J.M., 1993, 
 ApJ 411, 674

\bibitem[Toscano et al.\ 1999a]{toscano}
 Toscano M., Britton M.C., Manchester R.N., Bailes M., Sandhu J.S., 1999a,
 ApJ 523, L171

\bibitem[Toscano et al.\ 1999b]{toscano1}
 Toscano M., Sandhu J.S., Bailes M., Manchester R.N., Britton M.C.,
 Kulkarni S.R., Anderson S.B., Stappers B.W., 1999b,
 MNRAS 307, 925

\bibitem[van den Heuvel\ 1994]{van}
 van den Heuvel E.P.J., 1994,
 A\&A 291, L39
 
\end{thebibliography}
\end{document}